\documentclass[prl,aps,twocolumn]{revtex4}
\usepackage{epsfig}
\usepackage{graphics}
\usepackage{hyperref}
\usepackage{amsmath}
\begin{document}

\title{
Huge Positive Magnetoresistance in Antiferromagnetic 
Double Perovskite Metals}

\author{Viveka Nand Singh and Pinaki Majumdar}

\affiliation{Harish-Chandra  Research Institute,
 Chhatnag Road, Jhusi, Allahabad 211019, India}

\date{5 July 2011}

\begin{abstract}
Metals with large positive magnetoresistance  are rare. We demonstrate that 
antiferromagnetic  metallic states, as have been predicted for the double 
perovskites, are excellent candidates for huge positive magnetoresistance. 
An applied field suppresses long range antiferromagnetic order leading to
a state with short range antiferromagnetic correlations that generate  
strong electronic scattering. The field induced resistance ratio can be more 
than {\it tenfold}, at moderate field, in a structurally ordered system, and 
continues to be almost twofold even in systems with $\sim 25 \%$ antisite 
disorder. Although our explicit demonstration is in the context of a two 
dimensional spin-fermion model of the double perovskites, the mechanism we 
uncover is far more general, complementary to the colossal negative 
magnetoresistance process, and would operate in other local moment metals
that show a field driven suppression of non-ferromagnetic order.  
\end{abstract}

\maketitle

There has been intense focus over the last two decades on
magnetic materials which display large negative 
magnetoresistance (MR) \cite{mr1,mr2,mr3}.
In these systems, typically, an applied magnetic field
reduces the spin disorder 
leading to a suppression of the resistivity. The field
may even drive an insulator-metal transition leading
to `colossal' MR \cite{mr1}.
Large {\it positive} MR is rarer, and seems counterintuitive
since an applied field should reduce magnetic disorder
and enhance conductivity.
We illustrate a situation in double perovskite \cite{dp1}
metals where an applied field can lead to enormous positive
MR. The underlying principle suggests that
local moment antiferromagnetic metals 
\cite{af-met1,af-met2,af-met3,af-met4}, at strong coupling, 
should in general be good candidates for such unusual field response.

The double perovskites (DP), A$_2$BB'O$_6$, 
have been explored \cite{dp1} mainly
as ferromagnetic metals (FMM) or antiferromagnetic insulators
(AFMI). The 
prominent example among the former is Sr$_2$FeMoO$_6$
(SFMO) \cite{dp2}, while a
typical example of the later is Sr$_2$FeWO$_6$. The  
FMM have moderate negative MR \cite{dp2,expt-asd-huang,expt-asd-nav}.
Like in other correlated oxides \cite{dag-rev}, 
the magnetic order in the
ground state is expected to be sensitive to electron doping
and it has been suggested \cite{dp-af-th1,dp-af-th2,dp-af-th3} 
that the FMM can give way to
an AFM metal (AFMM) on increasing electron density.
The AFM order is driven by electron delocalisation and,
typically,  has lower spatial symmetry than the parent
structure. 
The conduction path in the AFM background is 
`low dimensional' and easily disrupted.

There is ongoing effort \cite{af-expt1,af-expt2}, 
but no clear indication yet
of the occurence of an AFMM on electron doping 
the FMM double perovskite. The problems are twofold: 
(a)~{\it Antisite defects:}  
in a material like SFMO,
substituting La for Sr to achieve a higher electron
density in the Fe-Mo subsytem tends to make the
Fe and Mo ionic sizes more similar (due to resultant
valence change), increases
the likelihood of 
% mislocation, {\it i.e}, 
antisite disorder (ASD), and suppresses magnetic order. 
(b)~{\it Detection:} 
even if an AFM state
is achieved, confirming the magnetic order is not possible
without neutron scattering. The zero field 
resistivity is unfortunately
 quite similar \cite{vn-pm-afm} to that of the FMM.

We discovered a remarkably simple indicator of an
AFMM state.  Using a real space Monte Carlo method 
on a two dimensional (2D) model of DP's 
we observe that:
(i)~For temperatures below the zero field $T_c$ of the
AFMM, a modest magnetic field can {\it enhance} the resistivity
{\it more than tenfold}. (ii)~The magnetoresistance is
suppressed by structural defects (antisite disorder)
 but is still
almost twofold for $\sim 25 \%$ mislocation of
B sites. (iii)~The enhancement in resistivity 
is related to the presence of short range correlated
AFM domains that survive the field suppression of global 
AFM order
and lead to enhanced electron scattering.
The mechanism suggests a generalisation to 3D, where
direct simulations are difficult, and to other local 
moment AFMM
systems where an applied field can suppress long range 
AFM order.

We first describe the structurally ordered DP model and will
generalise it to the disordered case later.
\begin{eqnarray}
H & =&\epsilon_{B}\sum_{i\in B}f_{i\sigma}^{\dagger}f_{i\sigma}+
\epsilon_{B'}\sum_{i\in B'}m_{i\sigma}^{\dagger}m_{i\sigma}
% -\mu {\hat N}  \cr
 -t\sum_{<ij>\sigma}f_{i\sigma}^{\dagger}m_{j\sigma} \cr
 &&~~ + J\sum_{i\in B} {\bf S}_{i}  \cdot
f_{i\alpha}^{\dagger}\vec{\sigma}_{\alpha\beta}f_{i\beta}
-\mu N -h \sum_i S_{iz}
\nonumber
\end{eqnarray}
$f$ is the electron operator on the magnetic, B,
site and $m$ is the operator on the non-magnetic, B',
site. $\epsilon_B$ and
$\epsilon_{B'}$ are onsite energies,
at the B and B' sites respectively.
The B and B' alternate along each cartesian axes.
$\Delta =\epsilon_{B} - \epsilon_{B'} $
is a  `charge transfer' energy.
$t$ is the hopping between nearest neighbour (NN)
 B and B' ions on the
square lattice and we will set $t=1$ as the reference scale.
$ {\bf S}_i$ is
the core B spin on the site ${\bf R}_i$.
We assume 
$S = \vert {\bf S}_i \vert =1$ and absorb the magnitude of the spin in the Hund's
coupling $J$ at the B site. We use $J/t \gg 1$.
When the `up spin' core levels
are fully filled, as for Fe in SFMO, the conduction electron is forced
to be {\it antiparallel} to the core spin. We have used $J>0$ to
model this situation. We have set the
{\it effective } level difference  $\Delta_{eff}=
(\epsilon_B -J/2) - \epsilon_{B'} =0$.
The chemical potential $\mu$ is chosen so that the electron
density is in the window for A type (stripe like) order
of core spins. $N$
is the total electron count and $h$
is the applied magnetic field in the ${\hat z}$ direction.
We have ignored next neighbour hopping and 
orbital degeneracy in the present model.  
We also focus on the 2D case here, where simulation
and visualisation is easier, and comment on the 3D 
case at the end.

The model has a FMM ground state 
at low electron density 
\cite{dp-af-th1,dp-af-th2,dp-af-th3}. This is 
followed by a phase with stripe-like
order - with FM  B lines coupled AFM 
in the transverse direction. We call this the `A type' phase. 
This is followed by a more traditional AFM phase where an up
spin B ion, say, is surrounded by four down spin B ions, and 
{\it vice versa} (the `G type' phase). 
We focus here on the A type phase since
it has a simple 3D counterpart and occurs at physically accessible
electron density 
% {\it Comment on the doping level..LaSFMO..}

We solve the spin-fermion problem via real space 
Monte Carlo using the traveling cluster method \cite{tca}.
This allows access to system size $32 \times 32$. 
We have used field cooling (FC) as well as zero field
cooling (ZFC) protocols. For ZFC the system is cooled to
the target temperature at $h/t=0$ and then a field is applied. 
We calculate the resistivity and the magnetic
structure factor peaks and also keep track of spatial
configurations of spins.
The optical conductivity is calculated via the 
the Kubo formulation, computing the
matrix elements of the current operator. The `dc conductivity'
is the low frequency average,   
$ \sigma_{dc} = 
(1/{\Delta \omega})
\int_{0}^{\Delta\omega}\sigma(\omega)d\omega $.
This is averaged over thermal configurations and disorder, as 
appropriate. Our `dc resistivity' $\rho$ is the inverse of this $\sigma_{dc}$.
$\Delta \omega = 0.05t$.

%-----------------------------------------------------------------------------
\begin{figure}[b]
\centerline{
\includegraphics[width=6.8cm,height=5.8cm,angle=0]{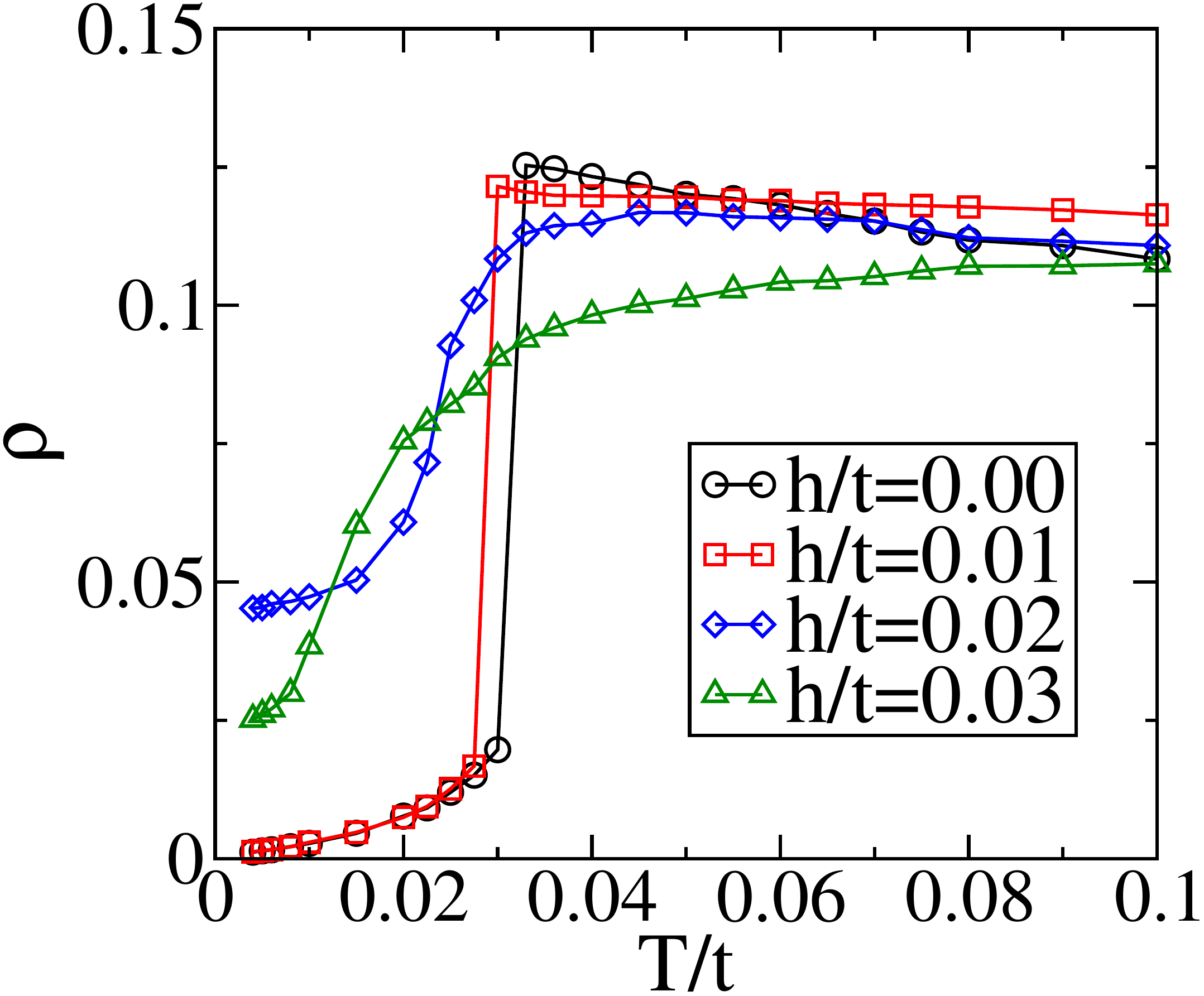}
}
\caption{Colour online:
The resistivity $\rho(T)$ in the absence of antisite disorder for
cooling in different applied fields. For temperatures below the zero field
transition, $T_c^0$, $\rho(T)$ increases on applying a field, and for
$T > T_c^0$ $\rho(T)$ decreases on applying a field. The ratio
$\rho(T,h)/\rho(T,0)$ can be very large as $T \rightarrow 0$ in this
field cooling situation. }
\end{figure}
%------------------------------------------------------------------------------
\begin{figure}[t]
\centerline{
\includegraphics[width=6.8cm,height=5.8cm,angle=0]{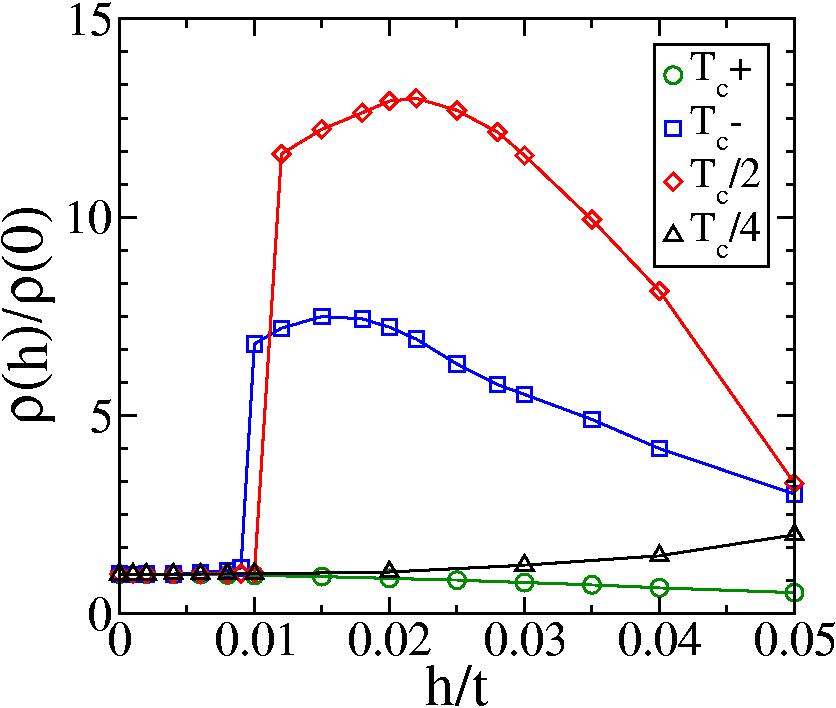}
}
\caption{Colour online: The field dependence of the resistivity
at different temperatures following a zero field cooling (ZFC)
protocol.
For $T = T_c^+$ (just above $T_c$) 
the zero field resistivity is already large and
$\rho$ 
decreases slightly with $h$ due to suppression of spin disorder.
For $T = T_c^-$ (just below $T_c$) 
and at $T_c/2$ there is a sharp increase
in resistivity at $h/t \sim 0.01$, with a peak around
$h/t \sim 0.02-0.03$ and a fall thereafter. This is consistent
with the trends seen in Fig.1. At $T=T_c/4$ this ZFC scheme does
not manage to create competing magnetic structures for
$h/t \sim 0.05$, possibly due to metastability of the parent
AFM pattern.
The field induced switching is therefore easiest achieved between
$T_c/2$ and $T_c$.
}
\end{figure}
%------------------------------------------------------------------------------

The A type pattern has {\it two possible orientations}
of the stripes,
either from bottom left to top right, or from bottom right to
top left. These are the two `diagonals' in 2D. The first corresponds 
to peaks in the structure factor $D({\bf q})$ at 
$\{{\bf Q}_{A1},{\bf Q}_{A2} \}$,
and the second to peaks at $\{{\bf Q}_{A3},{\bf Q}_{A4}\}$. For reference,
${\bf Q}_{A1} = \{\pi/2,\pi/2\}$, 
${\bf Q}_{A2} = \{3\pi/2,3\pi/2\}$, 
${\bf Q}_{A3} = \{\pi/2,3\pi/2\}$, 
${\bf Q}_{A4} = \{3\pi/2,\pi/2\}$. 
The ferromagnetic peaks in $S({\bf q})$ are 
at ${\bf Q}_{F1}= \{0,0\}$ and ${\bf Q}_{F2} = \{\pi,\pi\}$.
The ordered configurations lead to peaks at {\it two} wavevectors
since the model has both magnetic and non-magnetic sites and
our wavevectors are defined on the overall B-B' lattice.

Let us first examine the FC results in resistivity, Fig.1.
Cooling at
$h/t=0$ leads to a sharp drop in resistivity at $T=T_c^0 \sim 0.032$,
where $T_c^0$ is the zero field transition temperature, and
$ \rho(T) \rightarrow 0$ as $T \rightarrow 0$.
Cooling at  $h/t =0.01$
leads to a small suppression in $T_c$
but the trend in $\rho(T)$ remains similar  to $h/t=0$.
Between $h/t=0.01$ and $h/t=0.02$, however, there is a
drastic change in $\rho(T)$, and, as we will see later,
in the magnetic state.
The primary effect is a sharp increase in the $T < T_c^0$
resistivity, with the $T \rightarrow 0$ resistivity now
being almost $40\%$ of the paramagnetic value. Even at
this stage it is clear that $\rho(T,h)/\rho(T,0)$ can
be very large as $ T \rightarrow 0$ and is $\sim 4$ for $T \sim T_c^0$
and $h/t=0.02$.
Increasing the field even further leads to a reduction in $\rho(T)$
over most of the temperature window since the field promotes a 
ferromagnetic state suppressing the AFM fluctuations. 

We can study the impact of the magnetic field also within
the ZFC scheme. Fig.2 shows the result of applying a field
after the cooling the system to four different temperatures,
(i)~$T_c^+$ (slightly above $T_c^0$), (ii)~$T_c^-$ (slightly below $T_c^0$),
(iii)~$T_c^0/2$ and (iv)~$T_c^0/4$.

For $T > T_c^0$, the applied field mainly suppresses
the AFM fluctuations, leading to a gradual fall in the
resistivity. This is weak negative MR. Below $T_c^0$ (where there
is already noticeable AFM order) and
at $T_c^0/2$ the resistivity remains almost unchanged till 
some value $h_c(T)$ and then switches dramatically. There is
a peak in the ratio $\rho(T,h)/\rho(T,0)$ at $h \gtrsim
0.02$ and then a gradual decrease. At $T_c^0/2$ the
ratio reaches a maximum  $\sim 12$.
At lower temperature, $T_c^0/4$, the field appears to have a 
much weaker effect, mainly because the update mechanism that 
we adopt does not allow a cooperative switching of the AFM
state at low $T$ till very large fields. 
Overall, there is a window 
of $T$ over which a moderate magnetic field can lead to
a several fold rise in resistivity.

A first understanding of the rise in resistivity can be
obtained from the magnetic snapshots of the system at
$T=T_c^0/2$ in Fig.3. We plot the correlation $f_i
= {\bf S}_0.{\bf S}_i$ in an equilibrium magnetic
snapshot, where ${\bf S}_0$ is a reference spin (bottom
left corner) and ${\bf S}_i$ is the spin at site
${\bf R}_i$. The left panel is at $h/t=0$ and shows a 
high level of A type correlation (stripe like pattern).
This is a `low dimensional' electron system since the
electron propagation is along the one dimensional stripes
in this 2D system. The stripe pattern has a high
degree of order so the scattering effects and resistivity 
are low. The second panel is at $h/t=0.008$, just below
field induced destruction of AFM order, and the pattern
is virtually indistinguishable from that in the first
panel.

%------------------------------------------------------------------------------
\begin{figure}[t]
\centerline{
\includegraphics[width=2.2cm,height=8.5cm,angle=270]{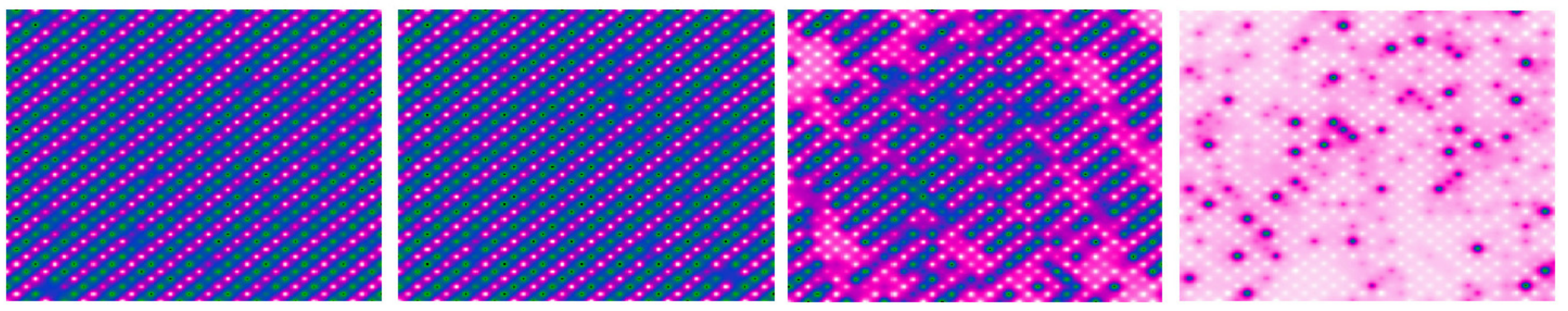}
}
\caption{Colour online: Evolution of spin correlations in the
clean system in response
to a magnetic field.  The plot shows
$f_i
= {\bf S}_0.{\bf S}_i$ in  magnetic
snapshots at different $h$, 
where ${\bf S}_0$ is a reference spin (bottom
left corner). $T=T_c^0/2$ and the fields are, from left to
right, $h/t=0,~0.008,~0.02,~0.05$. }
\end{figure}
%------------------------------------------------------------------------------

The third panel in Fig.3 is at $h/t=0.02$ where the applied
field has suppressed {\it long range} A type order. 
However, there are strong A type fluctuations that
persist in the system and they lead to a pattern of
short range ordered A type patches with competing
orientations, $\{ {\bf Q}_{A1}, {\bf Q}_{A2}\}$ and
$\{ {\bf Q}_{A3}, {\bf Q}_{A4} \}$, in a spin polarised
background. This patchwork leads to a high resistivity,
higher than that in leftmost panel, since the ferromagnetic
paths are fragmented by intervening A type regions,
while the A type regions are poorly conducting due
to their opposite handedness.
In the last panel the field, $h/t=0.05$,  is large enough so 
that even the AFM fluctuations are wiped out and the
spin background is a 2D ferromagnet with extremely short
range inhomogeneties. The resistance here
is significantly below the peak value.

Let us summarise the physical picture that emerges in the
non disordered system before analysing the effect of
antisite disorder.
The ingredients of the large MR are the following:
(i)~An AFM metallic phase, without too much quenched
disorder so that the resistivity in the magnetically 
ordered state is
small. (ii)~Field induced suppression of the AFM 
order at $h=h_c(T)$, say, replacing the ordered state 
with AFM correlated spins in a FM background,
leading to a high resistivity state.
In contrast to the standard negative MR scenario, where
an applied field pushes the system from a spin disordered
state to a spin ordered state, here the applied field 
pushes the ordered (AFM) state towards spin disorder.
%------------------------------------------------------------------------------
\begin{figure}[b]
\centerline{
\includegraphics[width=8.5cm,height=8.0cm,angle=0]{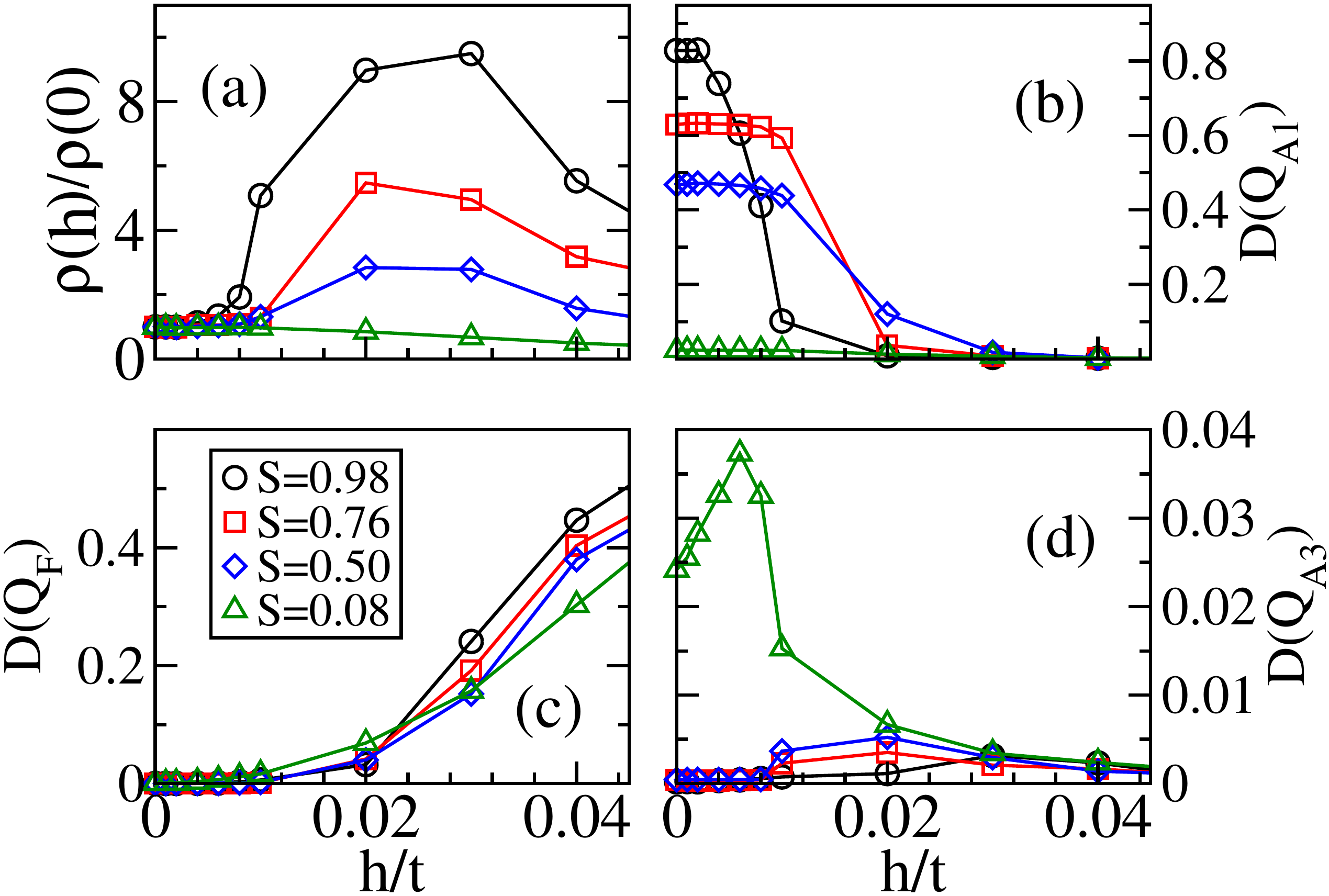}
}
\caption{Colour online: Field response in the presence of antisite
disorder. The temperature is $T=T_c^0/2$, where $T_c^0$ is the
$T_c$ at $h/t=0$ in the non disordered system. The structural
order parameters are indicated in panel (c), $S = 1-2x$ where $x$ is
the fraction of mislocated sites. The legend is same
for all the panels. The results are obtained via ZFC.
Panel (a)~Field dependence of resistivity, normalised to 
$\rho$ at $h/t=0$.
(b)~Magnetic structure factor at the major AFM peak ${\bf Q}_{A1}$.
The value is same at ${\bf Q}_{A2}$ also. (c)~Growth in the
FM structure factor with $h$. (d)~Growth in the complementary AFM
peak ${\bf Q}_{A3}$, the result is same for ${\bf Q}_{A4}$.
The primary observation here is the continuation of the
clean limit results, with suppressed magnitude, down to $S=0.50$.  }
\end{figure}
%---------------------------------------------------------
While our concrete results are 
in the case of a 2D `double perovskite' model and a stripe-like
ground state, the principle above is far more general and 
should apply
to other non ferromagnetic ordered states in two or three
dimensions and to microscopic models that are very
different from the double perovskites. We will discuss this issue at the
end.

Defects are inevitable in any system and in particular one expects
antisite disorder  in the DP's. The concentration of such defects
may actually 
increase on electron doping a material like SFMO,
due to the valence change,
and we need to check if the large MR is wiped out
by weak disorder.
The presence of ASD affects the zero field
magnetic state itself, as we have discussed elsewhere \cite{vn-pm-afm},
and the field response has to be understood with reference
to this $h/t=0$ state. 
Let us first define the augmented model
to describe the presence~of~ASD.
 
%------------------------------------------------------------------------------
\begin{figure}[t]
\centerline{
\includegraphics[width=7.5cm,height=10.0cm,angle=0]{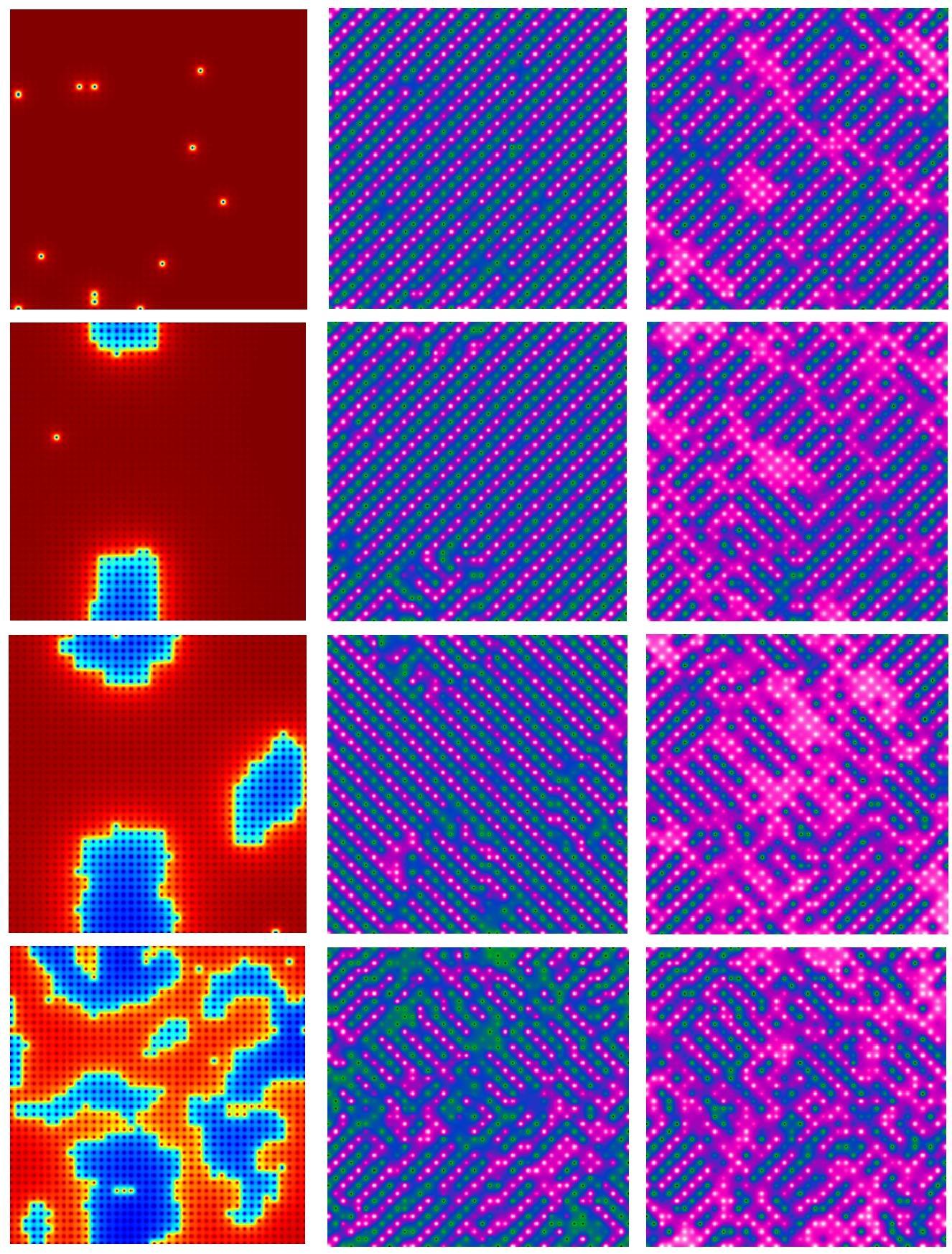}
}
\caption{Colour online:
Field response in the antisite disordered systems at $T= T_c^0/2$, for
the $S$ values in Fig.4.
The left panels indicate the structural domains  in
the antisite disordered systems. The middle column  shows the
spin correlations at $h/t=0$, note that rows 1-3 show significant
A type order, while the pattern in the 4th row has AFM domains
of both orientations and so suppressed long range order. The right 
column shows the spin correlations at $h/t=0.02$. In rows 1-3 the
AFM pattern gets fragmented and FM regions show up.
In row 4 the finite field pattern is not significantly
different from the $h/t=0$ case. Overall, the field enhancement
of `spin disorder' is large in the first three cases
but modest at strong ASD.
 }
\end{figure}
%------------------------------------------------------------------------------

We generalise the clean model  by including a configuration 
variable $\eta_i$, with
$\eta_i =1$ for B sites and $\eta_i =0$ for B' sites.
In the structurally ordered DP the $\eta_i$
alternate along each axis. We consider progressively
`disordered' configurations, generated through an annealing process
\cite{ps-pm-asd}.
These mimic the structural domain pattern observed 
\cite{asd-tok-dom,asd-dd-dom} in the real DP.
For any specified $\{\eta\}$ background the electronic-magnetic model
has the form:
$$
H= H_{loc} \{ \eta\} + H_{kin} \{ \eta\} + H_{mag} \{ \eta\}
$$
$H_{loc} \{ \eta\}~=~\epsilon_{B}
\sum_{i}\eta_{i}f_{i\sigma}^{\dagger}f_{i\sigma}+\epsilon_{B'}\sum_{i}\left(1-\eta_{i}\right)
m_{i\sigma}^{\dagger}m_{i\sigma} $, 
the hopping term connects NN sites,
irrespective of whether they are B or B':
$H_{kin} \{ \eta\} = 
-t_1\sum_{{\langle i,j \rangle} \sigma}\eta_i \eta_j
f_{i\sigma}^{\dagger}f_{j\sigma} $
$ -t_2\sum_{{\langle i,j \rangle} \sigma}
(1-\eta_i) (1-\eta_j)
m_{i\sigma}^{\dagger}m_{j\sigma} $
$ -t_3\sum_{{ \langle i,j \rangle} \sigma}
( \eta_i+\eta_j-2\eta_i \eta_j)
( f_{i\sigma}^{\dagger}m_{j\sigma}+h.c)$.
The magnetic interactions include the Hund's coupling on
B sites, and AFM superexchange $J_{AF} = 0.08t$  between two NN B sites:
$H_{mag} \{ \eta\} = 
J\sum_{i}\eta_{i}{\bf S}_{i}.f_{i\alpha}^{\dagger}{\vec \sigma}_{\alpha\beta}f_{i\beta} 
+J_{AF}\sum_{\left\langle i,j\right\rangle }\eta_{i}
\eta_{j} {\bf S}_i\cdot{\bf S}_j$.
For simplicity we set the NN hopping amplitudes 
$t_1=t_2=t_3=t$.

We use the simplest indicator $S=1-2x$ (not to be confused with
the magnitude of the core spin), to characterise
these configurations, where
$x$ is the fraction of
B (or B') atoms that are  on the wrong sublattice.
Fig.4 shows the resistivity ratio $\rho(h)/\rho(0)$ 
at $T= T_c^0/2$, in panel (a), and the field dependence
of structure factor peaks in panels (b)-(d).
Fig.5 first column shows the structural motifs on
which the magnetism is studied.

Down to $S=0.50$ the ratio $\rho(h)/\rho(0)$ has a behaviour similar 
to the clean case, Fig.2, but the peak ratio reduces to $\sim 3$
for $S=0.50$. There is a corresponding field induced 
suppression in the
principal AFM peak ${\bf Q}_{A1}$ and an enhancement of the
FM peak ${\bf Q}_F$. The complementary AF peak ${\bf Q}_{A3}$
slowly increases with $h$, has a maximum around $h/t=0.02$
(where the disconnected AFM domains exist) and falls at large
$h$ as the system becomes FM overall. The trend that we had
observed in the clean limit is seen to survive to significant
disorder. At $S=0.08$, where the B-B' order is virtually
destroyed, the $h/t=0$ state, Fig.5 last row, has no long 
range AFM order. It is already a high resistivity state and
an applied field actually leads to weak negative MR.

Let us place our results in the general context of AFM metals.
(i)~Theory: we are aware of one earlier effort 
\cite{af-met-th} in calculating the MR of AFM metals (and semiconductors),
assuming electrons weakly coupled to an independently ordering local moment
system. Indeed, the authors suggested that AFM {\it semiconductors} could
show positive MR.
Our framework focuses on field induced suppression of long range AFM order,
rather than perturbative modification, and the positive MR shows up even
in a `high density' electron system. The electron-spin coupling is also
(very) large, $J/t \gg 1$,  and cannot be handled within Born
scattering.

(ii)~Experiments: 
the intense activity on oxides has led to discovery of a few AFM 
metals, {\it e.g}, in the manganite La$_{0.46}$Sr$_{0.54}$MnO$_3$ 
\cite{af-met1}, in the ruthenate 
Ca$_3$Ru$_2$O$_7$ \cite{af-met2,af-met3}, and CaCrO$_3$ \cite{af-met4}.
Of these Ca$_3$Ru$_2$O$_7$ indeed shows large increase in resistivity 
with field induced growth of FM order \cite{af-met3}. For the manganite
and CaCrO$_3$ (where the resistivity is too large) we are not aware of
MR results across the field driven transition. Some heavy fermion metals too
have AFM character 
but the field driven transition affects the local moment itself, making
the present scenario inapplicable.

(iii)~Generalisation:  
the information about the spin configurations at any $(T,h)$
is encoded in the structure factor, $D({\bf q})$. Let us put down a 
form for $D({\bf q})$ and suggest how it affects the resistivity.
To simplify notation we will assume that the AFM peak is at
one wavevector ${\bf Q}$, while the FM peak is at $\{0,0\}$.
The AFM phase has an order parameter $m_{AF}$, say, while,
for $h > h_c(T)$, there is induced FM order of magnitude
$m_F$. 
Assuming that the dominant {\it fluctuations} in the
relevant part of the $(T,h)$ phase diagram 
are at ${\bf q} \sim {\bf Q}$, we can write:
$D({\bf q}) \sim m^2_{AF} \delta({\bf q} - {\bf Q}) 
+ A/(1 + ({\bf q} - {\bf Q})^2 \xi^{2})$ when  
$h < h_c(T)$, and 
$D({\bf q}) \sim m^2_{F} \delta({\bf q})
+ A'/(1 + ({\bf q} - {\bf Q})^2 \xi^{2})$ when $h > h_c(T)$.
$m_{AF}, m_F$ and $\xi$ depend on $(h,T)$, $A$ depends on $m_{AF}$ 
and $\xi$
and $A'$ on $m_F$ and $\xi$.
The amplitudes $A$ and $A'$ vanish as the corresponding $m$
tend to saturate (since the spins get perfectly ordered).
The  delta functions  $D({\bf q})$ dictate the bandstructure while
electron scattering is controlled by the Lorentzian part.

Consider three cases (a)~$T =T_c^+,~h=0$, (b)~$T = T_c^-,~h=0$, and
(c)~$T = T_c^-,~h > h_c(T)$. 
In (a) there is no order, so A is large, and $\rho = \rho_a$, say.
For (b) even if $\xi$ were the same as in (a), the presence of a large 
order parameter would suppress $A$ and hence the scattering. We call
this $\rho = \rho_b << \rho_a$ (assuming there is indeed a large
order parameter and no significant background resistivity due
to impurities). 
If we apply a field such that $m_{AF} \rightarrow 0$ but $m_F$ is
{\it still small}, then the structure factor crudely mimics the
paramagnetic case, and we should have $\rho_c \approx \rho_a$.
If all this is true, then just beyond field suppression of AFM
order (and for $T$ just below $T_c$) we should get
$\rho_c/\rho_b \gg 1$.
Broadly, if the appearance of AFM order with reducing $T$ 
leads to a sharp drop in $\rho$ then the field induced 
resistivity ratio can be large. This is independent of
dimensionality and microscopic detail.
A caution: as $ T \rightarrow 0$, the applied field would
drive a first order transition from a large $m_{AF}$ state
to one with large $m_F$ and  weak scattering. The scenario above
will not work, as our Fig.2 illustrated.

%(iv)~In a model with quantum spins, will the $T =0$ resistivity 
%vanish in the clean limit, unlike in Fig.1?

%(v)~Details: put overall scale of resistivity: ${\hbar a_0}/{\pi e^2} $. Scale
%for $h$: if $t=0.3$eV and $S=5/2$, then $h/t=0.01$ is
%$\sim 10$~Tesla. The normalised $n$ in LaSFMO and the $n$ for
%the AFMM in this paper.

{\it Conclusion:}
We have studied the magnetoresistance in an antiferromagnetic metal
motivated by the prediction of such a phase in the double perovskites.
Beyond the modest field needed for suppression of long range
antiferromagnetic order, the system shows almost tenfold increase
in resistivity near $T_c$. The effect originates from strong 
antiferromagnetic fluctuations in the field induced 
ferromagnetic background.  The large positive magnetoresistance, 
though suppressed gradually, survives the presence of significant 
antisite disorder.
The principle that we uncover 
behind this `colossal positive magnetoresistance' 
should be applicable to other local moment based AFM metals as well.
  
{\it Acknowledgments:}
We acknowledge use of the Beowulf Cluster at HRI, and thank
S. Datta and G. V. Pai for discussions.
PM acknowledges support from a DAE-SRC Outstanding 
Research Investigator Award, and the DST India (Athena).

\end{document}